\pdfoutput=1
\newcommand*{\ATLASLATEXPATH}{}
\documentclass[cernpreprint, atlasdraft=false, texlive=2016, USenglish]{atlasdoc}
\usepackage{\ATLASLATEXPATH atlaspackage}
\usepackage{\ATLASLATEXPATH atlasbiblatex}
\usepackage{\ATLASLATEXPATH atlasphysics}

\usepackage{xcolor}
\usepackage{lscape}
\usepackage{subfig}
\usepackage{multirow}

\newcommand\twb{\ensuremath{tWb}\xspace}
\newcommand\powheg{\textsc{Powheg-Box-v2}}
\newcommand\powhegRes{\textsc{Powheg-Box-Res}}

\AtlasTitle{Probing the quantum interference between singly and doubly resonant top-quark production in $pp$ collisions at $\sqrt{s}=13$ TeV with the ATLAS detector}
\author{ATLAS Collaboration}
\AtlasAbstract{%
This Letter presents a normalized differential cross-section measurement in a fiducial phase-space region where interference effects between top-quark pair production and associated production of a single top quark with a $W$ boson and a $b$-quark are significant.
Events with exactly two leptons ($ee$, $\mu\mu$, or $e\mu$) and two $b$-tagged jets that satisfy a multi-particle invariant mass requirement are selected from $36.1$ fb$^{-1}$ of proton--proton collision data taken at $\sqrt{s}=13$ TeV with the ATLAS detector at the LHC in 2015 and 2016.
The results are compared with predictions from simulations using various strategies for the interference. 
The standard prescriptions for interference modeling are significantly different from each other but are within $2\sigma$ of the data.
State-of-the-art predictions that naturally incorporate interference effects provide the best description of the data in the measured region of phase space most sensitive to these effects.
These results provide an important constraint on interference models and will guide future model development and tuning.
}
\AtlasJournal{Phys.\ Rev.\ Lett.}
\PreprintIdNumber{CERN-EP-2018-087}
\AtlasJournalRef{Phys.\ Rev.\ Lett. 121 (2018) 152002}
\AtlasDOI{10.1103/PhysRevLett.121.152002}

\hypersetup{pdftitle={ATLAS document},pdfauthor={The ATLAS Collaboration}}

\begin{document}
\maketitle

Top-quark pair ($t\bar{t}$) production is one of the most widely studied processes at the Large Hadron Collider (LHC) and is a key background to many searches for physics beyond the Standard Model (BSM).  The differential cross-section for $t\bar{t}$ has been measured~\cite{%
  TOPQ-2016-01,TOPQ-2016-04,CMS-TOP-16-007,CMS-TOP-16-014,CMS-TOP-17-002}
 and calculated~\cite{Czakon:2011xx,Czakon:2015owf,Czakon:2016dgf}
across a wide kinematic range with high accuracy.  However, all of these results treat the decay of the top quark to a $b$-quark and $W$ boson in the narrow-width approximation, separating $t\bar{t}$ production from production of a single top quark in association with a $W$ boson and a $b$-quark (\twb).
Due to their identical $WWbb$ final states, processes with one or two timelike top-quark propagators (called singly and doubly resonant, respectively) interfere.
Standard ad hoc methods of modeling this interference~\cite{Frixione:2008yi,Hollik:2012rc,Demartin:2016axk,ATL-PHYS-PUB-2016-020} are a significant source of uncertainty for many BSM searches~\cite{Aaboud:2016lwz,SUSY-2016-17,bib:SUSY-2014-07,Aaboud:2016nwl,SUSY-2016-15,Aad:2016qqk}.
Traditional measurements of production of a single top quark with an associated $W$ boson ($tW$) are designed to be insensitive to such effects~\cite{TOPQ-2015-16,TOPQ-2016-12,CMS-TOP-17-018-tmp,
}.
Recent fixed-order calculations of the full next-to-leading-order (NLO) $pp\to \ell^{+}\nu\ell^{-}\bar{\nu} b\bar{b}$ process~\cite{Denner:2010jp,Denner:2012yc,Bevilacqua:2010qb,Cascioli:2013wga,Frederix:2013gra} include proper treatment of the interference and have set the stage for corresponding predictions
matched to a parton shower~\cite{Jezo:2016ujg}.
However, there are no measurements available to assess the modeling in a region sensitive to interference effects.

This Letter presents a novel way to test different models of the interference between $t\bar{t}$ and \twb, using 36.1 fb$^{-1}$ of proton--proton ($pp$) collision data at $\sqrt{s}=13$~\TeV\ collected with the ATLAS detector in 2015 and 2016.  
The measurement targets the dilepton final state, characterized by a pair of oppositely charged leptons ($ee$, $\mu\mu$, or $e\mu$) originating from $W$-boson decays,\footnote{Events involving $W\rightarrow \tau\nu$ decays with a subsequent decay of the $\tau$-lepton to either $e\nu_e\nu_\tau$ or $\mu\nu_\mu\nu_\tau$ are included in the signal.} associated with jets containing $b$-hadrons ($b$-jets) and missing transverse momentum due to undetected neutrinos.  
The contributions from doubly and singly resonant amplitudes (and hence also their interference) to the combined cross-section depend on the invariant mass of the $bW$ pairs in the event, $m_{bW}$.
In this analysis, the charged lepton is used as a proxy for the $W$ boson and a differential cross-section is measured 
as a function of the invariant mass of a $b$-jet and a lepton.
There is ambiguity in forming this mass, so

\begin{align*}
m_{b\ell}^\text{minimax} \equiv \text{min}\{\text{max}(m_{b_1\ell_1},m_{b_2\ell_2}),\text{max}(m_{b_1\ell_2},m_{b_2\ell_1})\}
\end{align*}

is used, where the $b_i$ and $\ell_i$ represent the two $b$-jets and leptons, respectively.
This choice is inspired by the minimax procedure used to construct the stransverse mass~\cite{Lester:1999tx,Barr:2003rg} and measure the top mass~\cite{CMS-TOP-11-027}.
At leading order, for doubly-resonant events at parton level, $m_{b\ell}^\text{minimax}<\sqrt{m_t^2-m_W^2}$, where $m_t$ and $m_W$ are the top-quark and $W$-boson masses, respectively.
Due to suppression of the doubly resonant contribution, the differential cross-section above this kinematic endpoint has increased sensitivity to interference effects.

ATLAS is a multipurpose particle detector designed with nearly full $4\pi$ coverage in solid angle~\cite{PERF-2007-01}.
Lepton and jet reconstruction and identification used in this paper are described in Ref.~\cite{SUSY-2016-16} and are briefly summarized in the following.  Electrons and muons are required to have transverse momentum $p_\text{T}>28$~\GeV, pseudorapidity\footnote{
Pseudorapidity is defined in terms of the angle $\theta$ with respect to the beam line as $\eta=-\ln\tan(\theta/2)$.
}
 $\eta$ satisfying $|\eta|<2.47$ (2.5) for electrons (muons), and meet a series of quality criteria~\cite{PERF-2016-01,PERF-2015-10}, denoted ``tight'' in Ref.~\cite{SUSY-2016-16}.  
Jets are clustered from topologically connected calorimeter cells~\cite{Aad:2016upy} using the anti-$k_t$ jet algorithm~\cite{Cacciari:2008gp} with radius parameter $R=0.4$ implemented in FastJet~\cite{Cacciari:2011ma} and calibrated to particle level~\cite{PERF-2016-04}.  
Jets are identified as originating from $b$-quarks with a multivariate classifier using observables sensitive to lifetimes, production mechanisms, and decay properties of $b$-hadrons~\cite{ATL-PHYS-PUB-2016-012}.
The tagging efficiency is determined in simulated \ttbar events to be 60\% (85\%) for the tight (loose) tagging criterion.

Samples of simulated data are used in the design of the measurement, estimation of the background, and the unfolding procedure. 
\textsc{Powheg-Box}~\cite{Alioli:2010xd} v1 and v2 were used to simulate $tW$ and \ttbar events, respectively, with \PYTHIA\ 6.428~\cite{Sjostrand:2006za}, the five-flavor scheme (5FS) CT10~\cite{Lai:2010vv} parton distribution function (PDF) set, and Perugia 2012~\cite{Skands:2010ak} collection of tuned parameters.
An identical configuration except using \PYTHIA~8.183 and \powheg\ for $tW$ was included for particle-level comparisons.
Alternative samples used \powheg\ or \MGMCatNLO\ (MG5\_aMC) 2.2.2~\cite{Alwall:2014hca}, each interfaced to Herwig++ 2.7.1~\cite{Bahr:2008pv} with the UE-EE-5 set of tuned parameters~\cite{Seymour:2013qka} and CT10 PDF set.
The $t\bar{t}+b\bar{b}$ process~\cite{Cascioli:2013era} was generated using \SHERPA\ 2.1.1~\cite{Gleisberg:2008ta}  plus OpenLoops~\cite{Cascioli:2011va} with the CT10 four-flavor scheme PDF.
The $V$+jets and $VV$+jets ($V=W,Z$) processes were generated with \SHERPA\ 2.2.1 and the CT10 PDF set.  Associated production of \ttbar\ with a boson ($t\bar{t}V$) was generated using MG5\_aMC 2.2.2 combined with \PYTHIA\ 8.186~\cite{Sjostrand:2007gs}, the NNPDF2.3LO PDF set~\cite{Ball:2012cx} and the A14 set of tuned parameters~\cite{ATL-PHYS-PUB-2014-021}.  All predictions, including the \ttbar and $tW$ processes, are normalized to next-to-next-to-leading-order or next-to-leading-order cross-sections~\cite{Catani:2009sm,Czakon:2011xx,Kidonakis:2010ux,Gleisberg:2008ta,Alwall:2014hca}.
All samples of simulated data were processed using the full ATLAS detector simulation~\cite{Aad:2010ah} based on \GEANT4~\cite{Agostinelli:2002hh}.

The signal process is combined $t\bar{t}+tWb$ production.\footnote{Although it also interferes with the signal process, the contribution from fully non-resonant $WWbb$ production is treated as background.  
Its contribution to the selected phase space is negligible compared to processes with top quarks.}
A calculation of the $e^\pm\nu\mu^\mp\bar{\nu} b\bar{b}$ process in the four-flavor scheme at NLO was implemented in \powhegRes~\cite{Jezo:2015aia,Jezo:2016ujg} with \PYTHIA\ 8.226.
Here, resonance-aware matching allows the inclusion of off-shell top-quark effects at NLO, and the interference term is included.
Alternatively, predictions are obtained from the exclusive \ttbar and $tWb$ samples described above, where the definition of the $tW$ process is chosen to enable combination with the corresponding \ttbar calculation.
This is nontrivial at NLO, where care must be taken to avoid double-counting $tWb$ events with $m_{bW}\sim m_t$.
The default scheme for combining the $t\bar{t}$ and $tW$ processes at NLO adopted here is diagram removal (DR)~\cite{Frixione:2008yi} in which all doubly resonant amplitudes are removed from the $tW$ sample.  
Other choices exist where doubly resonant contributions are canceled out by gauge-invariant subtraction terms (diagram subtraction, DS)~\cite{Frixione:2008yi} or are only included in the interference terms (DR2)~\cite{Hollik:2012rc,ATL-PHYS-PUB-2016-020}.
For a more detailed review of possible $tW$ definitions, see Ref.~\cite{Demartin:2016axk}.
Finally, all \ttbar events with $b$-jets not associated with top-quark decays are classified as \ttbar + heavy flavor (\ttbar+ HF) and treated separately from the signal process.

Events are selected with single-lepton triggers~\cite{TRIG-2016-01} and required to have a pair of opposite-charge leptons ($e^{\pm}e^{\mp}$, $\mu^{\pm}\mu^{\mp}$, $e^{\pm}\mu^{\mp}$). 
Events with a same-flavor lepton pair having invariant mass $m_{\ell\ell} <10$~\GeV\ or within 15~\GeV\ of the $Z$-boson mass are rejected to suppress contributions from low-mass resonances and $Z$+jets.
Events are required to have exactly two jets with $\pt>25$~\GeV\ and $|\eta|<2.5$ which satisfy the tight $b$-tagging criterion and no additional jets that pass the looser $b$-tagging requirement. 
This $b$-jet veto suppresses $t\bar{t}$+ HF events, which can have large $m_{b\ell}^\text{minimax}$ when a selected $b$-jet does not originate from a top-quark decay.

A combination of data-driven and simulation-based methods is used to estimate backgrounds to the $\ttbar+\twb$ signal process.
The dominant background at high $m_{b\ell}^\text{minimax}$ is \ttbar+ HF, where a $b$-jet from a top-quark decay is not identified. 
This contribution is estimated from data events with at least three jets that are $b$-tagged according to the tight criterion. 
Simulated data is used to extrapolate the \ttbar+ HF yield measured in this region to the two-$b$-tag signal selection, 
giving a prediction $1.49\pm0.05\text{(stat.)}\pm0.20\text{(syst.)}$ times larger than the prediction obtained using \textsc{Powheg}+\PYTHIA\ 6.
This is consistent with the results of previous measurements, finding scale factors from 1.1 to 1.7 depending on the selection criteria~\cite{TOPQ-2014-10,HIGG-2017-03,CMS-TOP-13-010,CMS-TOP-16-010,CMS-HIG-17-026}.
Figure~\ref{fig:recombl}(a) shows the $m_{b\ell}^\text{minimax}$ distribution for events passing the three-b-tag selection, constructed from the two b-jets with largest \pt.  The leading two b-jets are both found to originate from top decays in 60\% of simulated tt+HF events when $m_{b\ell}^\text{minimax}$ is below 160 GeV and less than 10\% when above.  Good agreement between data and prediction across the distribution demonstrates that the additional jet from heavy flavor is well-modeled.
The next largest background is from $Z+$jets production, which is estimated in an analogous manner 
from data events with same-flavor leptons satisfying an inverted $m_{\ell\ell}$ requirement.
In both cases, the \ttbar contribution is subtracted before estimating the scale factor.
Various checks show that this does not bias the measurement in the signal region phase space.
Finally, there is a small contribution from non-prompt and misidentified leptons arising from photon conversions,  heavy-flavor hadrons decaying leptonically, and jets misidentified as leptons.  Following Ref.~\cite{bib:TOPQ-2013-04},
this background is estimated using events with same-charge lepton pairs, after subtracting the prompt lepton contribution.
Minor contributions from $t\bar{t}V$ and $VV$+jets are estimated using simulation.
Uncertainties in the simulation-based extrapolations are described below.
The $t\bar{t}+\twb$ signal process accounts for 95\% of events passing the full selection,
with remaining background contributions subtracted from the data before unfolding the signal process to particle level.
In Figure~\ref{fig:recombl}(b), the data are compared to the predicted event yields for both the DR and DS schemes.

The unfolding procedure corrects detector-level\footnote{Detector level refers to the measured outputs of the detector; particle level refers to the particles which interact with the detector.} observables to particle level using a Bayesian method~\cite{DAgostini:1994zf} with one iteration, optimized to minimize the average uncertainty per bin.
The particle-level selection is defined to be as close as possible to the detector-level selection to minimize simulation-based corrections for acceptance effects and the detector resolution when unfolding.  The definitions of particle-level objects are given in Ref.~\cite{ATL-PHYS-PUB-2015-013} with the following choices and modifications: (1) jets are clustered from all simulated particles with a mean lifetime $\tau > 30 $~ps excluding muons and neutrinos to reduce model dependence, (2) jets are identified as $b$-jets if a $b$-hadron
is found within the jet cone. Particle-level events must pass the same event selection as detector-level events, including the $m_{\ell\ell}$ requirement. To avoid contamination from \ttbar+ HF production, events with three or more particle-level $b$-jets with $p_\text{T}>5$~\GeV\ are rejected.

\newcommand{\RecoPlotText}{The total predicted events and uncertainties are shown for both the DR and DS definition of the $tW$ process. }
\newcommand{\overflow}[1]{The rightmost bin of #1 distribution includes contributions from events beyond the displayed axis limit.
}
\begin{figure}[h!]
\centering
\subfloat[][]{\includegraphics[width=0.5\textwidth]{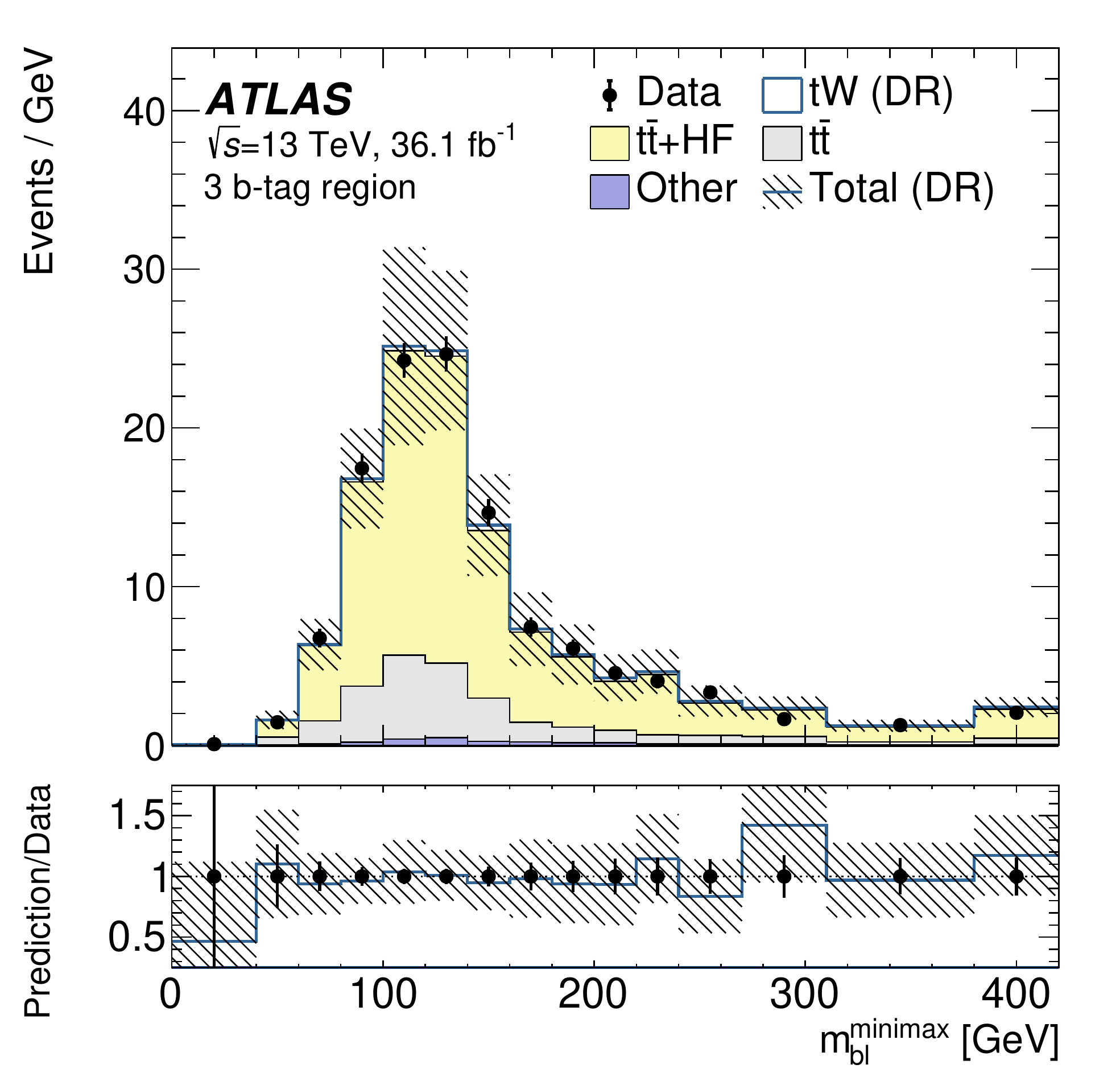}}%
\subfloat[][]{\includegraphics[width=0.5\textwidth]{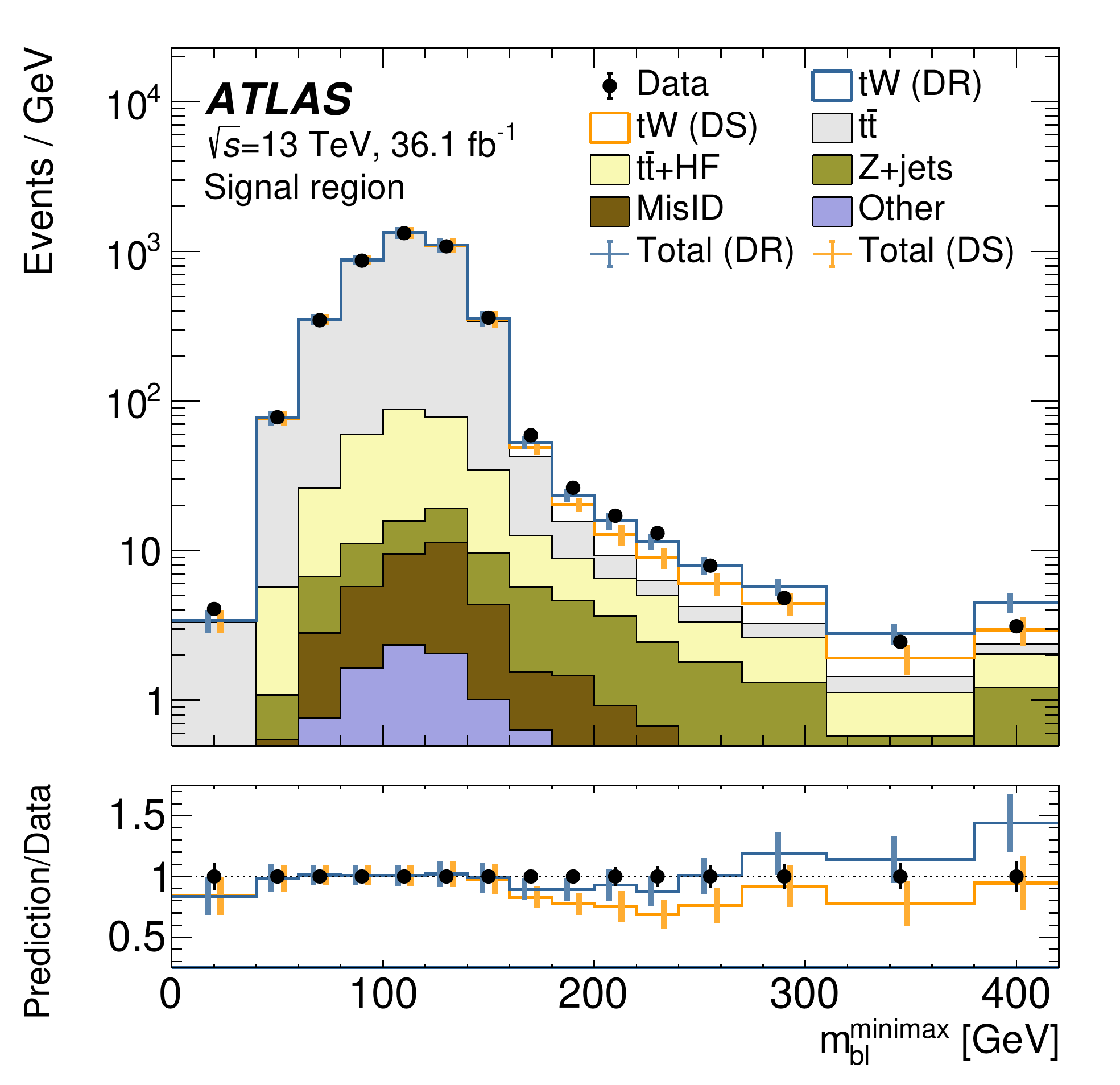}}
\caption{
  (a) The $m_{b\ell}^\text{minimax}$ distribution in the three-$b$-tag region, constructed from the two \bjets with largest \pt. 
  The predicted $t\bar{t}+$ HF contribution from simulation is scaled to match observed data in this region.
  The hashed band indicates the uncertainty on the total number of predicted events, where the DR scheme is used to estimate the minor contribution from the $tW$ process.
  (b) The detector-level $m_{b\ell}^\text{minimax}$ distribution, with signal selection and background estimation as described in the text. 
  The total predicted events are shown for both the DR and DS definitions of the $tW$ process, with uncertainties on the respective estimates indicated by separate error bars.
  Uncertainties include all statistical and systematic sources.
\overflow{each}
}
\label{fig:recombl}
\end{figure}

There are two categories of systematic uncertainties in the measurement: experimental and theoretical modeling.
These affect the result via the background prediction that is subtracted from data or through the model used to unfold the data to particle level.
Experimental uncertainties result from potential mismodeling in the reconstruction and identification of the jets~\cite{PERF-2016-04},
$b$-jets~\cite{PERF-2016-05}, and leptons~\cite{PERF-2016-01,PERF-2015-10}.  The background subtraction introduces uncertainty from the limited number of events in the control regions.  A suite of simulation samples with alternative settings are used to assess the theoretical uncertainties in modeling the \ttbar, $tW$, \ttbar+ HF, and $Z$+jets processes~\cite{ATL-PHYS-PUB-2016-004,ATL-PHYS-PUB-2016-003}.
A further uncertainty is assessed by varying the composition of the $\ttbar+tWb$ signal according to the uncertainty in the total cross-sections of the singly and doubly resonant processes.
An additional uncertainty is assessed for \ttbar+ HF by comparing the prediction obtained using \textsc{Powheg}+\PYTHIA6 with that using the \SHERPA \ttbar + \bbbar sample.
Furthermore, to ensure that the bias from the choice of interference scheme used in the unfolding is small, the procedure is repeated using the DS scheme.
Finally, as another test of the unfolding, the particle-level $m_{b\ell}^\text{minimax}$ spectrum is reweighted to attain better agreement between the corresponding detector-level distribution and the data.  Unfolding this reweighted distribution using the nominal unweighted simulation gives a measure of the method non-closure, which is assessed as an additional uncertainty~\cite{Malaescu:2009dm}.
The systematic uncertainty due to experimental sources ranges from 1\% to 14\%, with leading contributions from the jet energy scale and resolution and the $b$-tagging efficiency.
Theoretical uncertainties associated with the modeling of processes with top quarks are generally the most important and range from 1\% to 22\% of the unfolded yields.  The separate uncertainty due to the interference treatment is subdominant (22\% in the largest bin of $m_{b\ell}^\text{minimax}$, elsewhere 1\%--8\%), and everywhere much smaller than the raw difference between the DR and DS scheme predictions. The size of the dataset leads to statistical uncertainties of up to 20\%.

\begin{figure}[h!]
\centering
\includegraphics[width=0.80\textwidth]{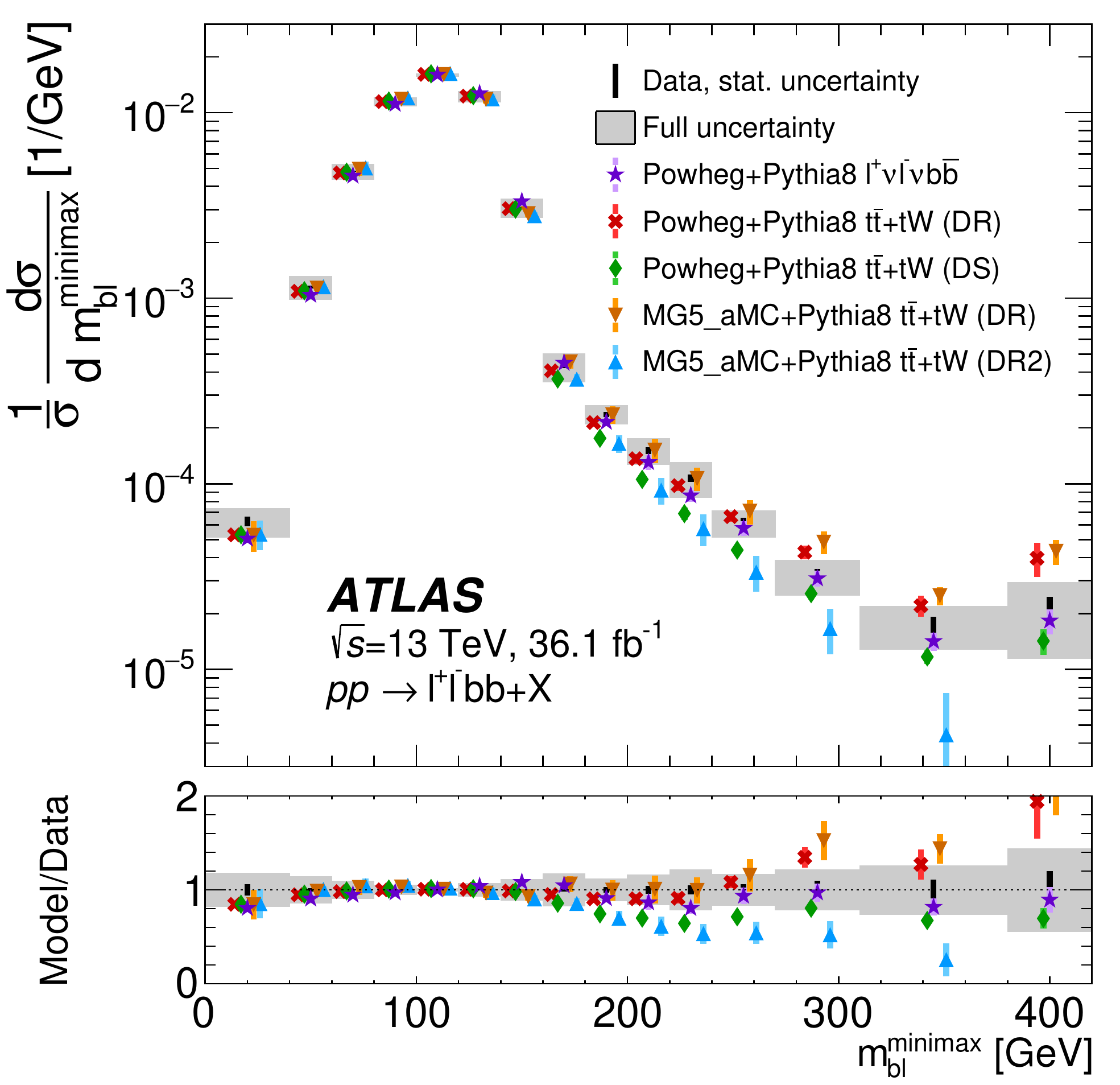}
\caption{The unfolded normalized differential $m_{b\ell}^\text{minimax}$ cross-section compared with theoretical models of the $\ttbar+\twb$ signal with various implementations of interference effects.
The uncertainty of each data point includes all statistical and systematic sources, while uncertainties for each of the MC predictions correspond to variations of the PDF set and renormalization and factorization scales.
\overflow{the}
}
\label{fig:results}
\end{figure}

Figure~\ref{fig:results} presents the differential cross-section observed in data, normalized
to the total observed cross-section with this selection.
Various predictions are also shown,
with uncertainties included from varying the PDF set~\cite{Butterworth:2015oua} and the renormalization and factorization scales.
A $\chi^2$ test statistic is constructed for the various models to assess the level of agreement with the data. Correlations among uncertainties of the unfolded distribution are included, as well as theory uncertainties on the signal predictions.
Results of the test are presented in Table~\ref{tab:pvals} as $p$-values, corresponding to the observed level of agreement over the full distribution as well as the subset $m_{b\ell}^\text{minimax}>160$~\GeV\ where the predicted differences due to interference are largest.

 \begin{table}[h!]
   \centering
   \caption{ $p$-values comparing data and predictions from events simulated with various models of the interference, all interfaced to \PYTHIA 8. Test statistics are constructed from the full $m_{b\ell}^\text{minimax}$ distribution and for the subset $m_{b\ell}^\text{minimax}>160$~\GeV.
 \label{tab:pvals}
   }
   \begin{tabular}{ l | l | c}
     \toprule
      \multirow{2}{*}{Model} & All & $m_{b\ell}^\text{minimax}$ \\ 
      & bins & $>160$~\GeV \\ \midrule     
	\POWHEGBOX $t\bar{t}+tW$ (DR)       & $0.71$           & $0.40$            \\
	\POWHEGBOX $t\bar{t}+tW$ (DS)       & $0.77$           & $0.56$           \\
     	MG5\_aMC $t\bar{t}+tW$ (DR)    & $0.14$  & $0.17$        \\
     	MG5\_aMC $t\bar{t}+tW$ (DR2)    & $0.02$  & $0.08$        \\
     	\POWHEGBOX $\ell^+\nu\ell^-\nu bb$  & $0.92$           & $0.95$            \\
    \bottomrule
   \end{tabular}
 \end{table}

The \twb prediction using the DR scheme gives a better description of the relative normalization of the region $m_{b\ell}^\text{minimax}\gtrsim m_t$ than the DS scheme.  However, the DS scheme better models the  $m_{b\ell}^\text{minimax}$ shape over the same range of values.
The DR and DS predictions generally bracket the data in the region of large $m_{b\ell}^\text{minimax}$, justifying the practice of applying their difference as a systematic uncertainty.
The DR2 scheme describes the data well up to the top-quark mass, but significantly underpredicts the data at higher masses\footnote{For this calculation, the effect of decaying the top quarks with Pythia instead of the default Madspin configuration can be up to 20\% at high $m_{b\ell}^\text{minimax}$.  However, this change leads to poorer agreement with data and the impact of using Madspin for DR2 is consistent with that seen for the corresponding DR prediction}.
The calculation from MG5\_aMC using the DR scheme is presented alongside the corresponding DR2 calculation to directly compare the two interference treatments with other inputs held constant.
The full $\ell^+\nu\ell^-\nu bb$ prediction\footnote{Generated $e\mu$ events are reweighted to account for events with same-flavor leptons and fully leptonic tau decays.} obtained from \powhegRes\ models $m_{b\ell}^\text{minimax}$ well across the full distribution, including the region beyond the top-quark mass where predictions using traditional models of the interference diverge.

\FloatBarrier
In summary, a measurement of a region sensitive to the interference between doubly and singly resonant top-quark pair production is presented.  
This is an original constraint on this interesting region of phase space that will be important for future model development and tuning.
The results are presented as a normalized fiducial differential cross-section, giving constraints on predictions for the full $t\bar{t}+\twb$ process.

\FloatBarrier
%-------------------------------------------------------------------------------
\section*{Acknowledgments}
%-------------------------------------------------------------------------------
% Version 14-Feb-2018
%-------------------------------------------------------------------------------
We thank CERN for the very successful operation of the LHC, as well as the
support staff from our institutions without whom ATLAS could not be
operated efficiently.

We acknowledge the support of ANPCyT, Argentina; YerPhI, Armenia; ARC, Australia; BMWFW and FWF, Austria; ANAS, Azerbaijan; SSTC, Belarus; CNPq and FAPESP, Brazil; NSERC, NRC and CFI, Canada; CERN; CONICYT, Chile; CAS, MOST and NSFC, China; COLCIENCIAS, Colombia; MSMT CR, MPO CR and VSC CR, Czech Republic; DNRF and DNSRC, Denmark; IN2P3-CNRS, CEA-DRF/IRFU, France; SRNSFG, Georgia; BMBF, HGF, and MPG, Germany; GSRT, Greece; RGC, Hong Kong SAR, China; ISF, I-CORE and Benoziyo Center, Israel; INFN, Italy; MEXT and JSPS, Japan; CNRST, Morocco; NWO, Netherlands; RCN, Norway; MNiSW and NCN, Poland; FCT, Portugal; MNE/IFA, Romania; MES of Russia and NRC KI, Russian Federation; JINR; MESTD, Serbia; MSSR, Slovakia; ARRS and MIZ\v{S}, Slovenia; DST/NRF, South Africa; MINECO, Spain; SRC and Wallenberg Foundation, Sweden; SERI, SNSF and Cantons of Bern and Geneva, Switzerland; MOST, Taiwan; TAEK, Turkey; STFC, United Kingdom; DOE and NSF, United States of America. In addition, individual groups and members have received support from BCKDF, the Canada Council, CANARIE, CRC, Compute Canada, FQRNT, and the Ontario Innovation Trust, Canada; EPLANET, ERC, ERDF, FP7, Horizon 2020 and Marie Sk{\l}odowska-Curie Actions, European Union; Investissements d'Avenir Labex and Idex, ANR, R{\'e}gion Auvergne and Fondation Partager le Savoir, France; DFG and AvH Foundation, Germany; Herakleitos, Thales and Aristeia programmes co-financed by EU-ESF and the Greek NSRF; BSF, GIF and Minerva, Israel; BRF, Norway; CERCA Programme Generalitat de Catalunya, Generalitat Valenciana, Spain; the Royal Society and Leverhulme Trust, United Kingdom.

The crucial computing support from all WLCG partners is acknowledged gratefully, in particular from CERN, the ATLAS Tier-1 facilities at TRIUMF (Canada), NDGF (Denmark, Norway, Sweden), CC-IN2P3 (France), KIT/GridKA (Germany), INFN-CNAF (Italy), NL-T1 (Netherlands), PIC (Spain), ASGC (Taiwan), RAL (UK) and BNL (USA), the Tier-2 facilities worldwide and large non-WLCG resource providers. Major contributors of computing resources are listed in Ref.~\cite{ATL-GEN-PUB-2016-002}.

\clearpage
\printbibliography

\clearpage
% ATLAS Collaboration author list
% Reference date of TOPQ-2017-05 is 2018-03-13
% Author list last updated on date 06-AUG-18
% Data extracted on 06-Aug-2018 for paper reference TOPQ-2017-05
% at 1:42pm
 
\begin{flushleft}
{\Large The ATLAS Collaboration}

\bigskip

M.~Aaboud$^\textrm{\scriptsize 34d}$,    
G.~Aad$^\textrm{\scriptsize 99}$,    
B.~Abbott$^\textrm{\scriptsize 124}$,    
O.~Abdinov$^\textrm{\scriptsize 13,*}$,    
B.~Abeloos$^\textrm{\scriptsize 128}$,    
D.K.~Abhayasinghe$^\textrm{\scriptsize 91}$,    
S.H.~Abidi$^\textrm{\scriptsize 164}$,    
O.S.~AbouZeid$^\textrm{\scriptsize 39}$,    
N.L.~Abraham$^\textrm{\scriptsize 153}$,    
H.~Abramowicz$^\textrm{\scriptsize 158}$
% ...
\bigskip
\\

% ...
$^{13}$Institute of Physics, Azerbaijan Academy of Sciences, Baku; Azerbaijan.\\
% ...
$^{34}$$^{(a)}$Facult\'e des Sciences Ain Chock, R\'eseau Universitaire de Physique des Hautes Energies - Universit\'e Hassan II, Casablanca;$^{(b)}$Centre National de l'Energie des Sciences Techniques Nucleaires (CNESTEN), Rabat;$^{(c)}$Facult\'e des Sciences Semlalia, Universit\'e Cadi Ayyad, LPHEA-Marrakech;$^{(d)}$Facult\'e des Sciences, Universit\'e Mohamed Premier and LPTPM, Oujda;$^{(e)}$Facult\'e des sciences, Universit\'e Mohammed V, Rabat; Morocco.\\
% ...
$^{39}$Niels Bohr Institute, University of Copenhagen, Copenhagen; Denmark.\\
% ...
$^{91}$Department of Physics, Royal Holloway University of London, Egham; United Kingdom.\\
% ...
$^{99}$CPPM, Aix-Marseille Universit\'e, CNRS/IN2P3, Marseille; France.\\
% ...
$^{124}$Homer L. Dodge Department of Physics and Astronomy, University of Oklahoma, Norman OK; United States of America.\\
% ...
$^{128}$LAL, Universit\'e Paris-Sud, CNRS/IN2P3, Universit\'e Paris-Saclay, Orsay; France.\\
% ...
$^{153}$Department of Physics and Astronomy, University of Sussex, Brighton; United Kingdom.\\
% ...
$^{158}$Raymond and Beverly Sackler School of Physics and Astronomy, Tel Aviv University, Tel Aviv; Israel.\\
% ...
$^{164}$Department of Physics, University of Toronto, Toronto ON; Canada.\\
% ...
$^{a}$ Also at Borough of Manhattan Community College, City University of New York, NY; United States of America.\\
$^{b}$ Also at Centre for High Performance Computing, CSIR Campus, Rosebank, Cape Town; South Africa.\\
$^{c}$ Also at CERN, Geneva; Switzerland.\\
$^{d}$ Also at CPPM, Aix-Marseille Universit\'e, CNRS/IN2P3, Marseille; France.\\
$^{e}$ Also at D\'epartement de Physique Nucl\'eaire et Corpusculaire, Universit\'e de Gen\`eve, Gen\`eve; Switzerland.\\
% ...
$^{*}$ Deceased

\end{flushleft}

% Created with Glance <Atlas.Glance@cern.ch>

%-------------------------------------------------------------------------------
% Auxiliary material - comment out the following line if you do not have any
%\include{TOPQ-2017-05-PAPER-auxmat}
%-------------------------------------------------------------------------------

\end{document}